\documentclass[aps,twocolumn, floatfix,superscriptaddress]{revtex4}
\usepackage{graphicx}
\usepackage{hyperref}
\usepackage{amsmath}
\usepackage{amsfonts}
\usepackage{mathtools,bm}
\usepackage[amssymb]{SIunits}
\usepackage{color}
\usepackage{mathrsfs}
\usepackage{animate}
\usepackage{soul}

\newcommand{\ket}[1]{|{#1}\rangle}

\begin{document}
\flushbottom

\title{Formation of nonlinear X-waves in condensed matter systems}

\author{David Colas}
\email{d.colas@uq.edu.au}
\affiliation{ARC Centre of Excellence in Future Low-Energy Electronics Technologies, School of Mathematics and Physics, University of Queensland, St Lucia, Queensland 4072, Australia}

\author{Fabrice P. Laussy}
\affiliation{Faculty of Science and Engineering, University of Wolverhampton, Wulfruna St, Wolverhampton WV1 1LY, United Kingdom}
\affiliation{Russian Quantum Center, Novaya 100, 143025 Skolkovo, Moscow Region, Russia}

\author{Matthew J. Davis}
\affiliation{ARC Centre of Excellence in Future Low-Energy Electronics Technologies, School of Mathematics and Physics, University of Queensland, St Lucia, Queensland 4072, Australia}

\begin{abstract}  
  X-waves are an example of a localized wave packet solution of the
  homogeneous wave equation, and can potentially arise in any area of
  physics relating to wave phenomena, such as acoustics,
  electromagnetism, or quantum mechanics.  They have been predicted in
  condensed matter systems such as atomic Bose-Einstein condensates in
  optical lattices, and were recently observed in exciton-polariton
  condensates.  Here we show that polariton X-waves result from an
  interference between two separating wave packets that arise from the
  combination of a locally hyperbolic dispersion relation and
  nonlinear interactions. We show that similar X-wave structures could
  also be observed in expanding spin-orbit coupled Bose-Einstein
  condensates.
\end{abstract}

 \date{\today} \maketitle

\section{Introduction}

X-waves are a well-known example of a localized wave packet, and have
been central to many efforts to generate optical pulses that are able
to resist diffraction~\cite{hernandez_book07a}.  They were originally
introduced as a superposition of  Bessel beams. that are non-diffracting solutions of
the homogeneous wave equation~\cite{lu92a}
\begin{equation}
  [\boldsymbol{\nabla}^2-(1/c^2)\partial^2_t]\Psi(\mathbf{r},t)=0,
\end{equation}
and can thus be encountered in a wide range of fields such as
acoustics, electromagnetism, quantum physics and potentially
seismology or gravitation.

Solitons and solitary waves are another famous type of non-spreading
wave packet which rely on a balance between dispersion and nonlinear
self-focussing to remain localized during
propagation~\cite{eiermann04a,amo09a,sich12a}.  However, X-waves do
not require any nonlinearity in the wave equation, a feature which they share with Bessel beams and other remarkable solutions, such as Airy beams---non-spreading solutions
of the Schr\"{o}dinger equation discovered by Berry and Balazs
\cite{berry79a} which have peculiar self-accelerating and self-healing
properties. 

X-waves, Bessel and Airy beams are non-physical
solutions since, like plane waves, they cannot be normalized and hence
would require an infinite energy to maintain their spectacular
properties through propagation. 
These solutions were thus initially considered a mathematical curiosity, but it was later realized and experimentally demonstrated that square-integrable approximations retain their surprising features for a significant amount of time~\cite{durnin87a,lu92b}. For Airy beams, such demonstration even came several decades later its original prediction~\cite{siviloglou07a,voloch13a}. A later experiment confirmed
Airy beams' self-healing property, showing their ability to self-reconstruct
even after strong perturbations, and also demonstrated their
robustness in adverse environments, such as in scattering and
turbulent media~\cite{broky08a}. Similarly, approximations of X-wave
packets must also reproduce their characteristic features, including
X-shape preserving propagation, but only for a finite time.

While Airy beams are typically produced by pulse shaping and
can be made arbitrarily close to their ideal (unphysical) blueprint,
it has been found that X-waves can conveniently be spontaneously
generated in dispersive and interacting media that feature a
hyperbolic dispersion, \textit{i.e.}, where the effective mass takes
opposite signs in transverse dimensions. In this instance they are
called ``nonlinear X-waves'' or X-wave
solitons~\cite{conti03a,trapani03a}. We will adopt this X-wave
terminology to refer to any similar phenomenology that results from
the combined effects of hyperbolic dispersion and interactions.  We
note that this is at best a finite-time approximation of an idealised
scenario which, as we shall discuss, opens new doors for alternative
interpretations in a realistic implementation.

X-waves were first discussed in a condensed matter context with a
theoretical proposal for their observation in an atomic Bose-Einstein
condensate (BEC)~\cite{conti04a}, where the hyperbolic dispersion can
be engineered by placing the BEC in a 1D optical lattice, ``bending''
the dispersion near the edge of the Brillouin zone.
Similar band engineering was proposed by Sedov
  \textit{et~al.} with Bragg
  exciton-polaritons~\cite{sedov15a}, using a periodical arrangement of
  quantum wells to realize hyperbolic metamaterials that support
  X-wave solutions.
  
However, a suitably hyperbolic
  dispersion naturally occurs with exciton-polaritons, which are
bosonic quasiparticles that arise from the strong coupling between
photons and excitons in semiconductors
microcavities~\cite{kavokin_book17a}. As a result of their hybrid
nature, they possess a highly non-parabolic and tunable dispersion
relation that provides inflection points, and thus regions of negative
effective mass, without the need for externally imposed potentials or Bragg polaritons. In 2D, one can find hyperbolic
regions that sustain X-waves solutions, as was first pointed out by
Voronych~\textit{et~al.}~\cite{voronych16a}, who also studied these
solutions extensively.  Another feature of polaritons is that their
interaction strength is tunable to some extent, either by changing the
excitonic (interacting) fraction or by altering the density of
particles, which allows the study of X-waves in both the weakly and
strongly interacting regimes.  Recently, the experimental observation
of polaritonic X-waves was reported~\cite{gianfrate18}.  In this
experiment polariton interactions were used to reshape an initial
Gaussian packet (easily created with a laser pulse) into an X-wave by
imparting it with a finite momentum above the inflection point of the
dispersion.  While this yielded a beautiful proof of principle of the
underlying idea, important questions remain open. In particular,
although one cannot hope to create an ideal X-wave, how close can one
can get through this interaction-based mechanism?  In a realistic
polariton system, how robust is the nonlinear instability that
converts a Gaussian wave packet into an X-wave~\cite{voronych16a}?
And for how long can an X-wave generated in this manner display its
expected characteristics?

To answer these questions, we examine the nonlinear X-wave formation
mechanism under the prism of the wavelet transform (WT), a spectral
decomposition that provides unique insights into the nontrivial
dynamics of wave packet propagation. Previously this technique has
been used to explain and fully characterize so-called self-interfering packets
(SIPs), another phenomenology observed with polaritons due to an
inflection point in the dispersion relation.  This results in
negative-mass effects (counter propagation) coexisting with normal
(forward) propagation, producing a constant flow of propagating
fringes~\cite{colas16a}. While purely a linear wave phenomenon, the
SIP can also be triggered due by a nonlinearity leading to the spread of the wave packet across the inflection point in momentum space.  The
formation of a SIP, powered by nonlinear interactions, was recently
observed in an atomic spin-orbit coupled
BEC~\cite{khamehchi17a,colas18a}.

In this paper, we show how the wavelet transform provides a new
understanding of the nature and formation of a nonlinear X-wave.  The
X-wave is indeed found to be a transient effect that occurs during the
reshaping of a Gaussian wave packet under the
combined effects of a non-parabolic dispersion and repulsive
interactions.  The spatial interference of two resulting sub-packets
travelling at different speeds accounts for the X-wave pattern.  The
polaritonic X-wave can thus be understood as another type of SIP
rather than a shape-preserving non-interacting ``soliton''. 
This confirms the self-interference mechanism is the key to
understanding the general problem of wave packet propagation under
nontrivial dispersion relations that feature inflection points and
thus both negative and infinite effective masses, either with or
without nonlinearity.

This paper is organized as follows.  In Sec.~\ref{sec:hyperbolic} we
introduce our method of analysis, and provide an idealized example of
X-wave formation in a complex wave equation with a purely hyperbolic
dispersion relation and a weak nonlinearity.  In
Sec.~\ref{sec:polaritons} we demonstrate how the same phenomenon
arises in the formation of X-waves in an exciton-polariton system.
Section~\ref{sec:SOCBEC} proposes how X-waves can be formed in atomic
Bose-Einstein condensates with artificial spin-orbit coupling, instead
of an additional optical lattice potential~\cite{conti03a}. We
conclude in Sec.~\ref{sec:conclusions}.

\section{Hyperbolic dispersion}
\label{sec:hyperbolic}
\begin{figure}[t!]
  \includegraphics[width=\linewidth]{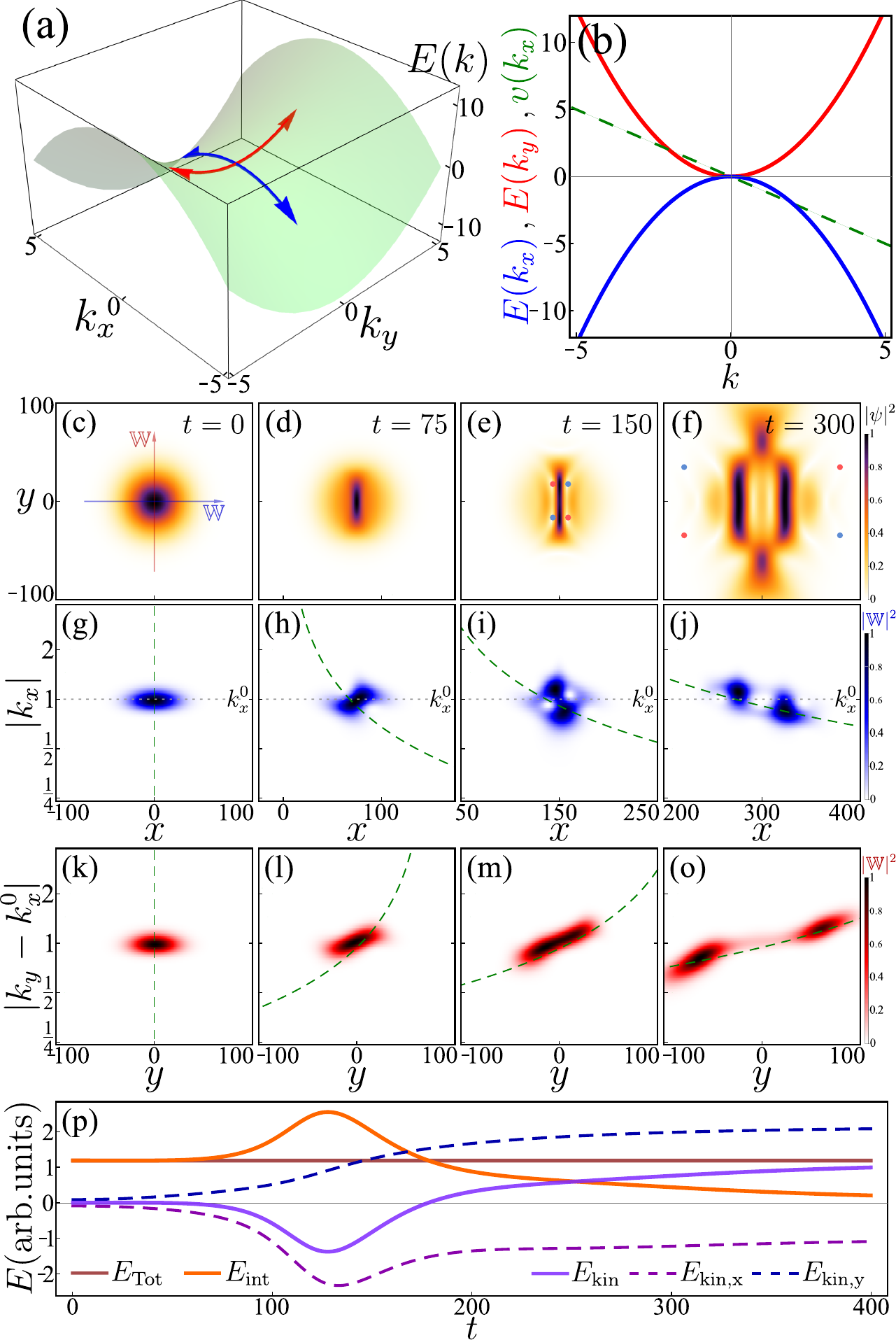}
  \caption{X-wave formation and propagation for a hyperbolic dispersion. (a) 2D
    hyperbolic-dispersion. (b) Effective dispersion along $k_x$ and
    $k_y$, with $v(k_x)$. (c--f) Evolution of the
    density $|\psi(x,y)|^2$ at selected times, starting from a
    Gaussian wave packet with  $\sigma= 20$ and imparted with a momentum
    $k_x^0=1$. Light-blue (red) dots indicate a $\pm 2\pi$ phase
    winding. (g--j) Corresponding wavelet energy density $|\mathbb{W}|^2$
    computed along the $x$ direction. (k--o) Idem but computed along the $y$
    direction. The green dashed curve shows the displacement of the
    $k_{x,y}$-components $d(k_{x,y})=v(k_{x,y})t$. (p) Evolution of the components of the energy: Total (brown), interaction (orange), kinetic
    (purple) with its two components along $x$ (dashed-dark purple) and
    $y$ (dashed-dark blue).  Parameters: $\hbar=m_x=-1$, $g=0.003$.  Supplemental Movie 1 provides an animation of the nonlinear X-wave formation for this system~\cite{footnote2}.} 
  \label{fig:1}
\end{figure}

We start with the simplest system allowing the generation of
nonlinear X-waves, a Gross-Pitaevskii equation for the field $\psi(x,y)$
\begin{equation}
i \hbar \partial_t \psi(x,y)=H_{\mathrm{hyp}}\psi(x,y).
\label{eq:GPE}
\end{equation}
The nonlinear operator
\begin{equation}
\label{eq:1}
H_\mathrm{hyp} = \frac{\hbar^2 k_x^2}{2 m_x} + \frac{\hbar^2 k_y^2}{2 m_y} + g|\psi(x,y)|^2\,,
\end{equation}
has masses of opposite signs in the $x$ and $y$ dimensions $m_x=-m_y$, and thus the system combines a hyperbolic dispersion with repulsive
interactions. A 3D representation of the hyperbolic dispersion
is shown in Fig.~\ref{fig:1}(a). The dispersion is parabolic in both
directions but with an inverted curvature in the $x$ direction, as
seen in Fig.~\ref{fig:1}(b). The last term in Eq.~(\ref{eq:1}) accounts
for the nonlinear interaction, characterised by the constant $g$. An
example of a nonlinear X-wave formation out of an initial Gaussian
wave packet imparted with a momentum $k_x^0$ is shown in Fig.~\ref{fig:1}(c--f)~\cite{footnote3}. One
can see the typical X-shape appearing in the density as
it propagates. Phase singularities with opposite winding also appear
when the X-wave fully forms, here marked as blue and red dots. However
the X-wave does not maintain its shape and breaks in larger packets at
long time, Fig.~\ref{fig:1}(f), much like square integrable Airy beam approximations
lose their self-accelerating property during
propagation~\cite{siviloglou07a}.

The X-wave formation mechanism can be better understood when
considering the field $\psi(\boldsymbol{r},t)$ in a different representation space. Various spectral representations of the wave function are accessible through the Fourier Transform, such as the space-energy $\psi(\boldsymbol{r},E)$ or the momentum-energy $\psi(\boldsymbol{k},E)$ (also called \textit{far-field}) representations. They can provide useful information on, \textit{e.g.}, relaxation processes yielding the Bose-Einstein condensation~\cite{estrecho18a}, or the characterization of topological effects with the presence of Dirac cones or flat-bands~\cite{jacqmin14a,baboux16a}.
\\
However, such representations are poorly adapted for the detection of a transient interference effect, as either the spatial or the temporal dynamics vanishes when integrating towards the momentum or energy domains. An alternative method of analysis is to make use of the Wavelet Transform (WT) --- a convenient manner in which to simultaneously represent the field in both position and momentum space at a given instant in time.

The WT was initially introduced in
signal processing to obtain a representation of the signal in both
time and frequency. It has proven to be particularly useful to analyse the interference
between different wave packets~\cite{baker12a} or more recently
the self-interference from a single wave
packet~\cite{colas16a,colas18a}. 
Unlike the usual Fourier Transform that is
based on the decomposition of the signal into a sum of unphysical states (delocalized
sine and cosine functions), the WT uses more physical states with localized wavelets
$\mathcal{G}$ as basis functions.

For a 1D wave packet $\psi(x)$, the general WT
reads~\cite{debnath_book15a}:
\begin{equation}
\mathbb{W}(x,k)=(1/\sqrt{|k|})\int_{-\infty}^{+\infty}\psi(x')\mathcal{G}^\ast [(x'-x)/k]\mathrm{d}x'\,.
\label{eq:5}
\end{equation}
A suitable representation when analysing Schr\"odinger wave
packets is the Gabor wavelet:
\begin{equation}
\mathcal{G}(x)=\sqrt[4]{\pi}\exp(i w_\mathcal{G} x)\exp(-x^2/2)\,,
\label{eq:6}
\end{equation}
This wavelet family consists of a Gaussian envelope, which is an
elementary constituent of the Schr\"{o}dinger dynamics, with an
internal phase that oscillates at a defined wavelet-frequency $\omega_\mathcal{G}$. The physical momentum~$k$ can be retrieved from the WT parameters (wavelet-frequency, grid specifics etc) using a numerical procedure that is detailed in Ref.~\cite{colas18a}. The quantity $|\mathbb{W}(x,k)|^2$ thus measures the cross-correlation between
the wavelet $\mathcal{G}(x)$ and the wave function
$\psi(x)$. This allows us to show in a transparent
way the position~$x$ in real-space of the different $k$-components of
the wave packet.

We apply the 1D-WT to the slice $\psi(x,y=0)$, \textit{i.e.}, along the
direction of propagation, and at different times of the X-wave
evolution, as shown in Fig.~\ref{fig:1}(g--j). The mechanism leading to
the X-wave formation appears clearly in this spectral
representation. At $t=0$, the wavelet energy density is tightly
distributed around the value $k_x^0$, Fig.~\ref{fig:1}(c,g), which is
the momentum initially imparted to the wave packet.  Since the wave packet is 
not spatially confined by any external potential, the initial interaction energy is converted into kinetic energy, leading to an
increase of the packet's spread in momentum space, as previously
observed in 1D systems~\cite{colas18a}. This first distortion can be
seen in the WT, Fig.~\ref{fig:1}(h), along with its consequence on the packet shape in real
space, which shrinks in the~$x$ direction,
Fig.~\ref{fig:1}(d). Indeed, in the direction of propagation, the
group velocity $v(k_x)=\partial_{k_x}E(k_x,0)$ decreases as the
momentum increases, see the dashed-green curve for $v(k_x)$ in
Fig~\ref{fig:1}(b). This means that a particle acquiring additional momentum
will travel more slowly. This feature is the key ingredient for the
X-wave formation. As the packet's spread in $k_x$ keeps increasing, the
latter effect leads to the break up of the initial packet into two
sub-packets, located at different $k_x$ and hence travelling at different
velocities. In Fig.~\ref{fig:1}(g--j), the green dashed line shows the
expected displacement of the $k_x$-components $d(k_x)=v(k_x)t$. In real space, the
sub-packet with the lowest momentum but with the highest group velocity
formed at the tail overtakes the other sub-packet
formed at a higher momentum but propagating at a lower velocity. The
spatial overlap of these two sub-packets creates the interference fringes that are at the heart of the peculiar X-shape of the wave packet.

We also apply the 1D-WT to the transverse direction of the center of
the packet while following its drift in $x$, \textit{i.e.}, we consider the
$y$-WT of $\psi(x=v(k_x^0)t,y)$.  The wavelet energy density
$|\mathbb{W}(y,k_y-k_x^0)|^2$~\cite{footnote4} is shown in
Fig.~\ref{fig:1}(k--o). The interactions also lead to an increase of
the packet's spread in $k_y$, followed by a breaking of the packet
into two distinct parts, but unlike for the $x$-direction, this
time the sub-packet with a higher momentum travels faster than the
one with a lower momentum, which prevents any interference from
occurring.

To complete the X-wave analysis, we take a closer look at the energy
exchanges occurring during the wave packet propagation. The Gaussian
wave packet set as an initial condition undergoes reshaping under the
joint action of the dispersion and repulsive interaction, under the
constraint of conservation of the total energy:
\begin{multline}
  \label{eq:sabfeb23172742GMT2019}
   E_\mathrm{Tot}=E_\mathrm{kin}+E_\mathrm{int} \\
   =\int\big[E(\mathbf{k})-E(\mathbf{k}_0)\big]|\psi(\mathbf{k})|^2\,d\mathbf{k}+\int\frac{g}{2}|\psi(\mathbf{r})|^4\,d\mathbf{r}\,.
\end{multline}
The kinetic energy $E_\mathrm{kin}$ is here computed in momentum space
in order to remove the important energy shift $E(\mathbf{k}_0)$
induced by the imparted momentum set in the initial condition. The
interaction energy $E_\mathrm{int}$ is more conveniently computed in
real space. The evolution of these different energy components is
shown in Fig.~\ref{fig:1}(p), with the total, interaction and kinetic
energies plotted in brown, orange and purple, respectively. It is also
instructive to consider the components of the kinetic energy
$E_{\mathrm{kin},x}$ and $E_{\mathrm{kin},y}$ along the $x$ and $y$
directions. They are plotted as dark purple and blue dashed lines,
respectively.  Note that at $t=0$, $E_\mathrm{kin}=0$ as
$E_{\mathrm{kin},x}=-E_{\mathrm{kin},y}$ since the initial packet is a
symmetrical Gaussian that spreads equally in both $x$ and $y$
directions of the hyperbolic dispersion with $E(k_x)=-E(k_y)$, which
cancels the overall kinetic energy. For the same reason, an increasing
spread in momentum along the $k_y$ direction leads to an increase of
$E_{\mathrm{kin},y}$ whereas an increasing spread in momentum along
the $k_x$ direction actually leads to a decrease of
$E_{\mathrm{kin},x}$. As the total energy has to be conserved, this
causes a momentary rise of the interaction energy as observed in
Fig.~\ref{fig:1}(p). The energy peak corresponds to the time of
maximum interference between the sub-packets, and also corresponds to
the time of the emergence of the phase singularities. At long times, when the new
packets spread out, all the interaction energy is converted into
kinetic energy, leaving the system behaving essentially as linear
waves. The above discussion illustrates neatly how the
  WT analysis captures the key physics that rules the wave
  packet reshaping, namely, the interplay between
  the hyperbolic dispersion and its resulting negative energy, and the
  interactions which peak to break the packet and create phase
  singularities.
\begin{figure}[t!]
  \includegraphics[width=\linewidth]{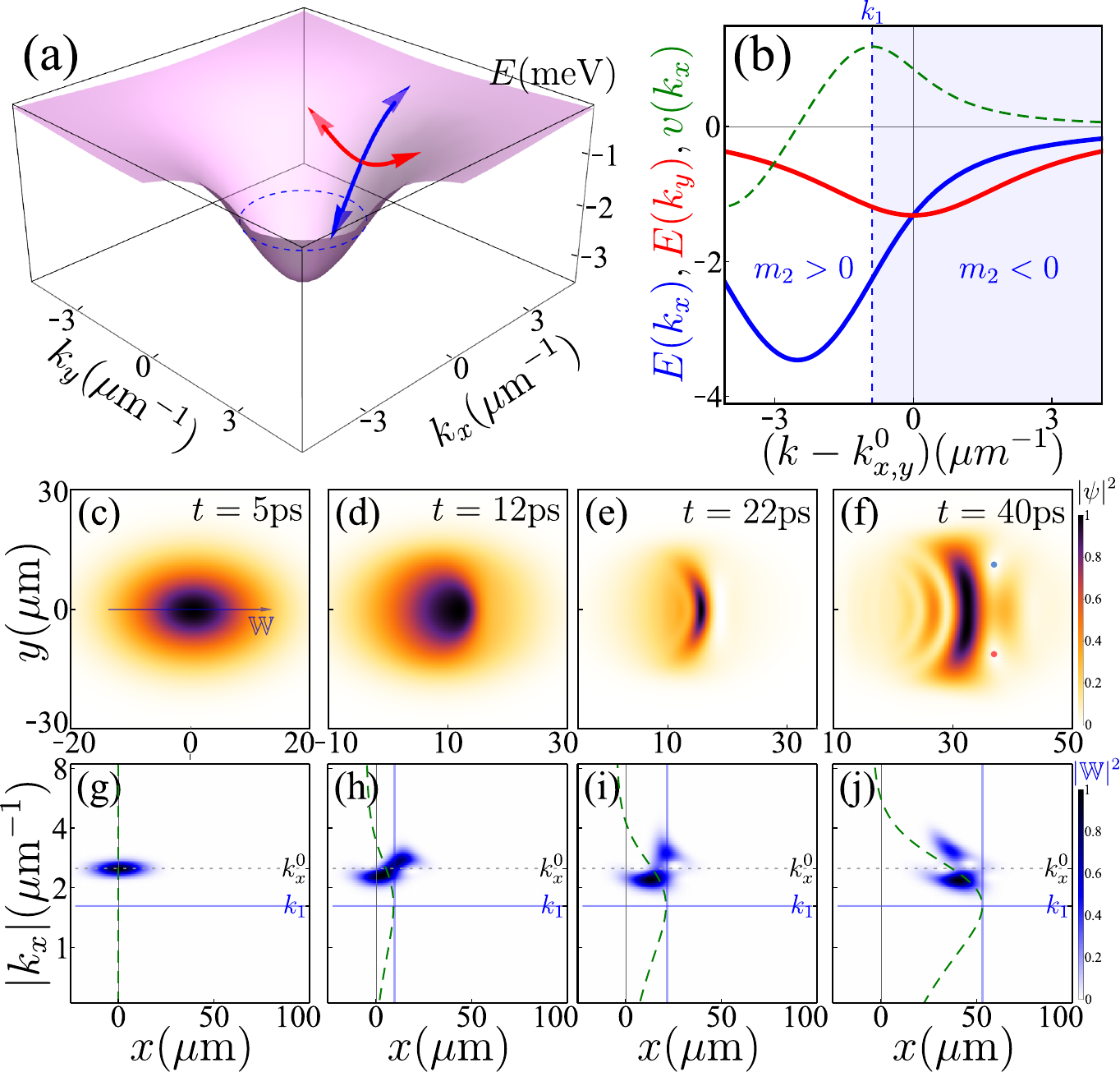}
  \caption{Exciton-polariton X-wave dynamics. (a) 2D lower polariton
    dispersion. The dashed-blue line indicates the position of the
    inflection points. (b) Effective dispersion in two transverse
    directions from the point
    $(k_x=\unit{2.5}{\mu\reciprocal\meter},k_y=\unit{0}{\mu\reciprocal\meter})$,
    with $v(k_x)$. (c--f) Evolution of the polariton density $|\psi|^2$
     at selected times. Light-blue (red) dots indicate a $\pm 2\pi$ phase
    winding. (g--j) Corresponding wavelet energy density $|\mathbb{W}|^2$
    computed along the $x$ direction. The packet is imparted with a momentum
    $k_x^0=\unit{2.5}{\mu\reciprocal\meter}$, above the inflection
    point $k_1$. The green dashed curve shows the displacement of
    the $k_x$-components $d(k_x)=v(k_x)t$. The vertical blue line correspond to
    displacement $d(k_1)$ delimiting the interference
    area.  Supplemental Movie 2 provides an animation of the nonlinear X-wave formation for this system~\cite{footnote2}.}
  \label{fig:2}
\end{figure}

\section{Exciton-polaritons}
\label{sec:polaritons}

We now study a realistic and physical exciton-polariton system, whose
dynamics can be well-captured by the following
two-component Gross-Pitaevkii operator~\cite{carusotto04a,gianfrate18}:
\begin{equation}
\label{eq:2}
H_{\mathrm{pol}}
= 
\begin{pmatrix}
 \frac{\hbar \boldsymbol{k}^2}{2 m_\mathrm{C}} + \Delta -i \frac{\gamma_\mathrm{C}}{2} & \frac{\Omega_\mathrm{R}}{2}\\ 
\frac{\Omega_\mathrm{R}}{2} & \frac{\hbar \boldsymbol{k}^2}{2 m_\mathrm{X}} -i \frac{\gamma_\mathrm{X}}{2} +g_\mathrm{X} |\psi_\mathrm{X}|^2
\end{pmatrix}\,,
\end{equation}
which acts on the spinor field
$\boldsymbol{\psi}=(\psi_\mathrm{C},\psi_\mathrm{X})^T$. The parameter $m_{\mathrm{C},(\mathrm{X})}$ is the photon (exciton) mass, $\Delta$ the detuning between the photonic and excitonic modes and $\Omega_\mathrm{R}$ their coupling strength. Both fields have an independent decay rate $\gamma_\mathrm{C,(X)}$. The nonlinearity is here introduced through the exciton-exciton interaction with a strength $g_\mathrm{X}$. Diagonalising the non-interacting and dissipationless part of the operator leads to  dressed upper and lower polariton modes:
\begin{multline}
  E_\mathrm{U,L} = \frac{\hbar \boldsymbol{k}^2}{2 m_{+}} +\frac{\Delta}{2}
  \pm \sqrt{\left(\frac{\hbar \boldsymbol{k}^2}{2 m_{-}}-\frac{\Delta}{2}\right)^2 +\left(\frac{\Omega_\mathrm{R}}{2}\right)^2 }\,,\\
\label{eq:3}
\end{multline}
where
$m_\pm = (m_\mathrm{C} \pm
m_\mathrm{X})/2m_\mathrm{C}m_\mathrm{X}$. In the following, we use a
similar set of parameters to Gianfrante \textit{et
  al.}~\cite{gianfrate18}. The lower branch $E_\mathrm{L}$ is plotted
in Fig.~\ref{fig:2}(a) and shows a circularly symmetric profile,
approximately parabolic at small $\boldsymbol{|k|}$, and possessing an inflection
point at $k_1=\unit{1.61}{\mu\reciprocal\meter}$ (dashed-blue
line). An X-wave can be generated by exciting the branch above the
inflection point in any given direction, where the effective
dispersion thus appears locally hyperbolic, as shown in
Fig.~\ref{fig:2}(b). The dynamical evolution of a polariton wave
packet can be obtained by solving the following equation:
\begin{equation}
i \hbar \partial_t \boldsymbol{\psi}=H_{\mathrm{pol}}\boldsymbol{\psi} + \mathbf{P}\,,
\label{eq:4}
\end{equation}
where
$\mathbf{P}= (\mathrm{LG}_{00}\mathrm{e}^{-(t-t_0)^2/2 \sigma_t}
\mathrm{e}^{-i \omega_\mathrm{L} t}\mathrm{e}^{-i k_x^0
  x},0)^\mathrm{T}$ stands for the pulse excitation. The photonic
field is excited with a Gaussian pulse arriving at time $t_0$, with a
temporal spread $\sigma_t$, an energy $\omega_\mathrm{L}$ and with an
imparted momentum $k_x^0$. The pulse parameters are chosen so that
only the lower branch is populated
($\omega_\mathrm{L}=\unit{-3}{\milli\electronvolt}$,
$\sigma_t=\unit{0.5}{\pico\second}$), preventing Rabi oscillations
between the two modes~\cite{dominici14a}. The initial momentum of the
pulse is set to be above the inflection point of the branch, at
$k_x^0=\unit{2.5}{\mu\reciprocal\meter}$. Selected time frames of the
density evolution are presented in
Fig.~\ref{fig:2}(c--f). Approximately $\unit{10}{\pico\second}$ after
the pulse arrival, the wave packet starts to distort, Fig.~\ref{fig:2}(d), then shrinks,
Fig.~\ref{fig:2}(e), before forming a typical X-shape profile Fig.~\ref{fig:2}(f) along with phase singularities. The formation of a vortex-antivortex pair is here again a consequence of the hyperbolic topology of the dispersion relation, which leads to an inwards polaritons flow along the propagation direction and outwards in the transverse one, as noted in Ref.~\cite{gianfrate18}.

The WT analysis reveals that the exact same formation mechanism
as for the ideal hyperbolic dispersion occurs in the polariton
system. Shortly after the pulse arrival, the wavelet energy density is
distributed around $k_x^0$, Fig.~\ref{fig:2}(g). The packet then
spreads in $k_x$ due to the interaction, Fig.~\ref{fig:2}(h), and narrows 
in the $x$-dimension in real space, Fig.~\ref{fig:2}(d). Above the inflection point
$k_1$, $v(k_x)=\partial_{k_x}E(k_x,0)$ decreases as the momentum
increases, which corresponds to the region where the effective mass
parameters $m_2=\hbar^2[\partial_k^2 E_\mathrm{L}(k)]^{-1}$ becomes
negative~\cite{colas16a}, see Fig.~\ref{fig:2}(b). The origin of the
subsequent X-wave formation is again identified as the result of an
interference between two sub-packets with different momenta and
travelling at different velocities, Fig.~\ref{fig:2}(g--j). The
observed X-wave profile slightly differs from the one obtained with
the symmetrically hyperbolic dispersion in Fig~\ref{fig:1}. This is
due to specifics of the polariton system, such as the asymmetry of
the branch above the inflection, which translates in a different
effective mass (in absolute value) in the transverse direction. Because the polariton system does not conserve the total energy, the analysis of the different energy components field is not as informative as it was for the hyperbolic case. 

Regardless of these relatively minor departures, it is clear that the
mechanism is otherwise the same as that discussed in the
previous section, which clarifies the nature and underlying formation mechanism
for the polaritonic nonlinear X-waves.

As a final remark in this section, we comment on the  the ``superluminal'' propagation of X-waves observed and discussed in Ref.~\cite{gianfrate18}.  Here, ``superluminal''
refers to the observed propagation of a density peak at a speed
exceeding  the speed of the packet's center-of-mass by $\sim6$\%. The
later speed is set by the initial imparted momentum $k_x^0$, \textit{i.e.},  the slope of the polariton dispersion at this point,
$$v(k_x^0)=\partial_{k_x} E_\mathrm{L} (k_x,0)\Bigr\rvert_{k_x=k_x^0}.$$
From the results presented in Fig.~\ref{fig:2}, we also
observe that the speed of the main peak exceeds the speed of the
center-of-mass by $\sim 5$--$6$\%. Note that the WT is here not a practical way to measure the peak velocity as it results in the decomposition of the two sub-packets at the origin of the interference peaks.
The simplest way to observe the superluminal propagation thus remains to track the position of the main peak in the real space density $|\psi(\mathbf{r},t)|^2$ and to find the corresponding velocity.

\begin{figure}[t!]
  \includegraphics[width=\linewidth]{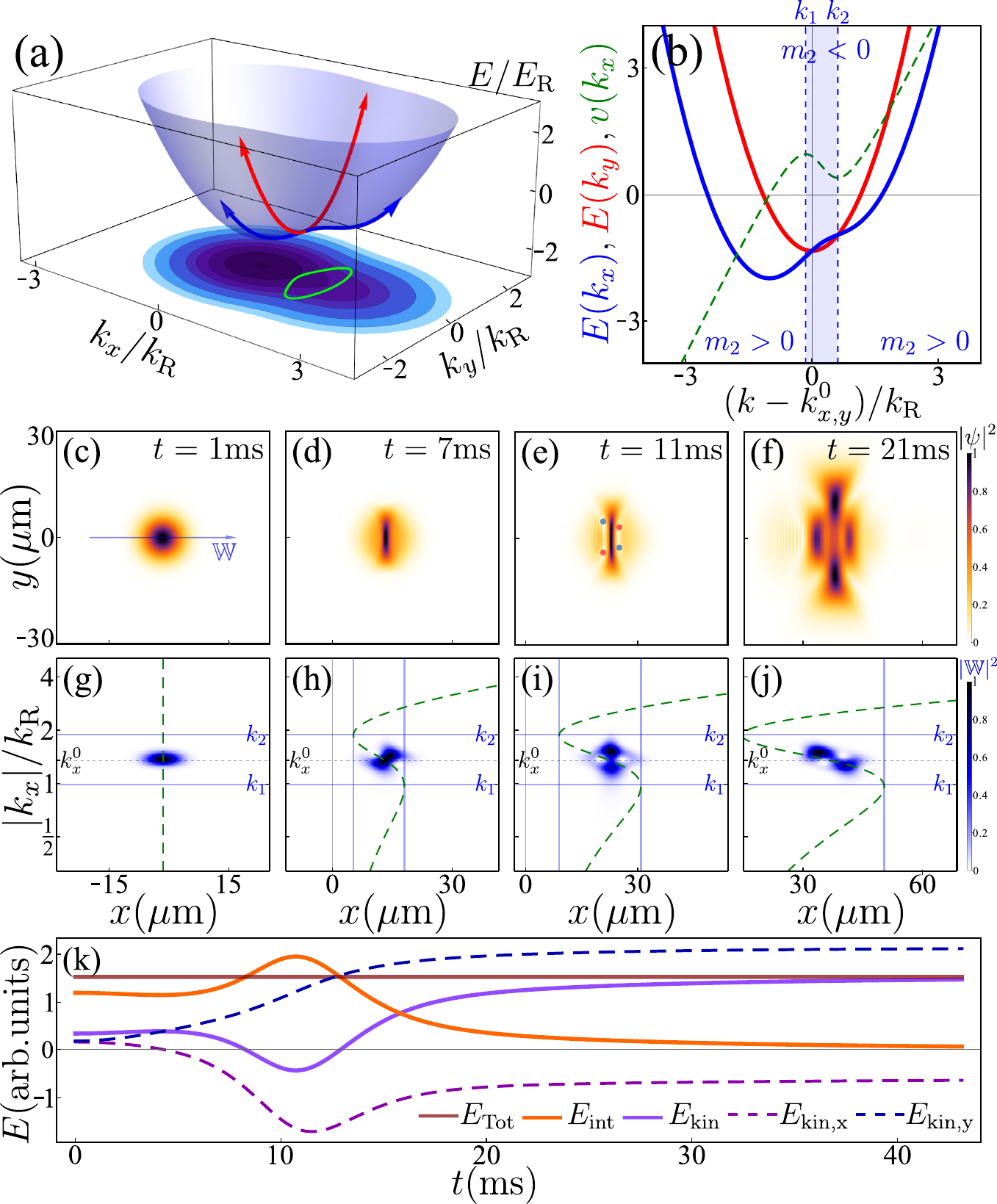}
  \caption{X-wave dynamics in a SOCBEC. (a) 2D-SOCBEC dispersion. The
    green line on the bottom projection encloses the inflection points
    region. (b) Effective dispersion in two transverse directions from
    the point $(k_x=1.35 k_\mathrm{R},k_y=0)$, with $v_x(k)$. (c--f)
    Evolution of the atomic density $|\psi|^2$ at selected
    times. Light-blue (red) dots indicate a $\pm 2\pi$ phase
    winding. We note that the vortices present in frame (e) have moved outside of the boundary of frame (f) (g--j) Corresponding wavelet energy density $|\mathbb{W}|^2$
    computed along the $x$ direction. The packet is
    imparted with a momentum $k_x^0$, between the inflection points
    $k_1$ and $k_2$. The green dashed curve shows the displacement of
    the $k_x$-components $d(k_x)=v(k_x)t$. The vertical blue lines correspond
    to displacements $d(k_1)$ and $d(k_2)$ delimiting the interference
    area. (k) Evolution of the different energies: total (brown),
    interaction (orange), kinetic (purple) with its two components
    along $x$ (dashed-dark purple) and y (dashed-dark blue).
    Supplemental Movie 3 provides an animation of the nonlinear X-wave formation for this system~\cite{footnote2}.}
  \label{fig:3}
\end{figure}
\section{Spin-orbit coupled Bose-Einstein condensates}
\label{sec:SOCBEC}
\begin{figure*}[t!]
  \includegraphics[width=\linewidth]{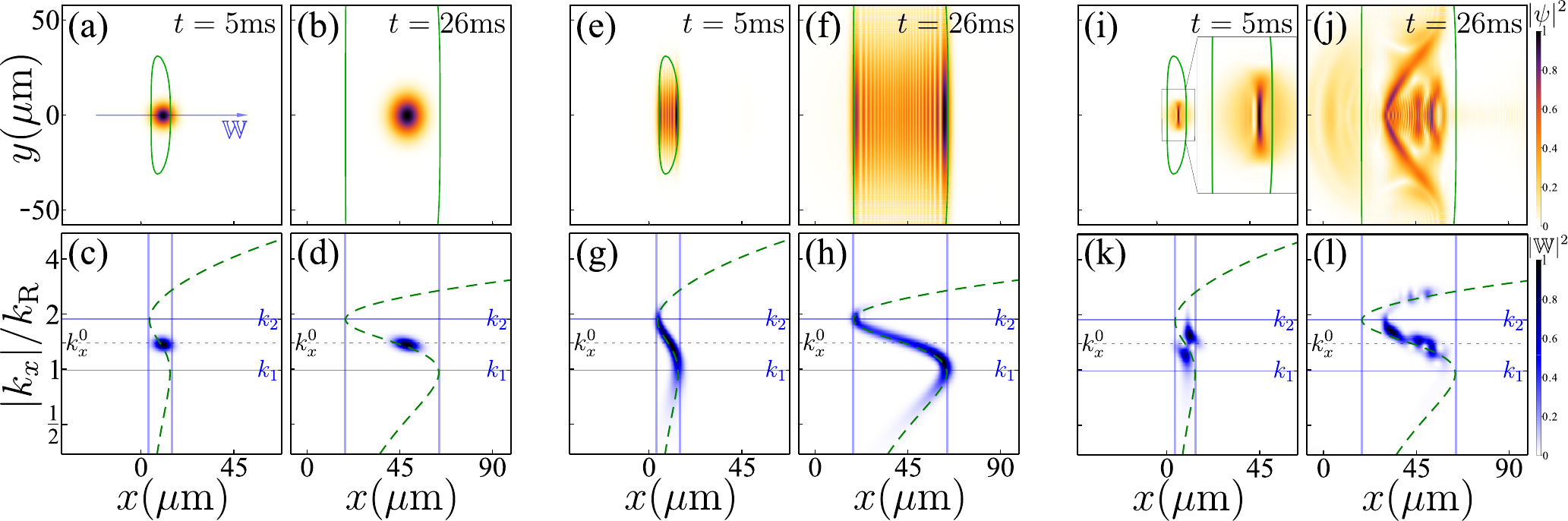}
  \caption{Wave packet propagation in 2D-SOCBEC. The top row shows the
    atomic density $|\psi(x,y)|^2$ at a given time of its
    evolution and the bottom row shows the corresponding wavelet
    energy density $|\mathbb{W}(x,k)|^2$, with a WT performed at
    $k_y=0$. The green line delimits the self-interference
    area. (a--d) Evolution from an initial Gaussian wave packet of width
    $\sigma_{\boldsymbol{r}}=\unit{3.5}{\micro\meter}$ in the linear
    regime ($g_\mathrm{2D}=0$). (e--h) Self-interfering regime,
    obtained from an initial Gaussian wave packet of width
    $\sigma_{\boldsymbol{r}}=\unit{0.35}{\micro\meter}$ in the linear
    regime ($g_\mathrm{2D}=0$). (i--l) The more strongly interacting regime compared to Fig.~\ref{fig:3} with $g_{2D}N=14\times10^{-4}E_\mathrm{R}$, from an initial Gaussian wave packet of
    width $\sigma_{\boldsymbol{r}}=\unit{3.5}{\micro\meter}$.}
  \label{fig:4}
\end{figure*}
We finally consider a third condensed-matter system in which SIPs have
been recently encountered in a one-dimensional setting --- a 1D-spin-orbit coupled
Bose-Einstein condensate (SOCBEC)~\cite{khamehchi17a,colas18a}. When
extended to two dimensions, this system also possesses the key elements to
generate nonlinear X-waves. 

A non-interacting 2D-SOCBEC can be described by the
following Hamiltonian~\cite{stanescu08a,linyj11a}:
\begin{equation}
\label{eq:7}
H_\mathrm{SOC}
= 
\begin{pmatrix}
\frac{\hbar (k_x^2 + k_y^2)}{2 m} +\gamma k_x +\frac{\delta}{2} & \frac{\Omega}{2}\\ 
 \frac{\Omega}{2} & \frac{\hbar (k_x^2 + k_y^2)}{2 m} -\gamma k_x -\frac{\delta}{2}
\end{pmatrix}\,,
\end{equation}
which acts on the spinor field
$\boldsymbol{\psi}=(\psi_\uparrow,\psi_\downarrow)^T$. Two hyperfine
pseudo-spin states up $\ket{\uparrow}=\ket{F=1,m_F=0}$ and down
$\ket{\downarrow}=\ket{F=1,m_F=-1}$ are coupled with the Raman
coupling strength $\Omega$ and detuned by $\delta/2$. We also
introduce $\gamma=\hbar k_\mathrm{R}/m $. The energy and momentum
units are set by $E_\mathrm{R}=(\hbar k_\mathrm{R})^2/2m$,
$E_\mathrm{R}$ and $k_\mathrm{R}$ being the recoil energy and the
Raman wavevector, respectively. 

Once
diagonalised, the individual dispersion relations of the two spin
states are mixed, leading to the upper ($+$) and lower ($-$) energy
bands:
\begin{equation}
\label{eq:8}
E_\pm (\boldsymbol{k}) = \frac{\hbar (k_x^2 + k_y^2)}{2m} \pm \sqrt{\left(\gamma k_x +\frac{\delta}{2}\right)^2 + \left(\frac{\Omega}{2}\right)^2}\, .
\end{equation}

The lower band $E_{-}(\boldsymbol{k})$ is plotted in
Fig.~\ref{fig:3}(a). Unlike the polariton dispersion, see
Fig~\ref{fig:2}(a), the 2D-SOCBEC dispersion is not circularly
symmetric and inflection points are only present in a finite region of
momentum space~\cite{footnote5}. This region can be determined
analytically. To do so, we make a change of coordinates
${k_x=k\cos(\theta),\,k_y=k\sin(\theta)}$ in Eq.~(\ref{eq:8}) to obtain
the dispersion relation $E(k,\theta)$ in polar coordinates. We can
then find the inflection points of the dispersion for each specific
angle $\theta$ by solving $\partial_k^2 E(k,\theta)=0$.  This yields
the following expression:
\begin{equation}
\label{eq:50}
k_{1,2}(\theta)=\frac{\delta}{4 k_\mathrm{R}}\pm \frac{\sec\theta}{4 k_\mathrm{R}}\sqrt{(2 k_\mathrm{R}\Omega \cos\theta)^{\frac{4}{3}}-\Omega^2}\,.
\end{equation}
These two solutions $k_{1,2}(\theta)$ are plotted as a light-green
line in Fig.~\ref{fig:3}(a) and form the delimiting region of momentum
space in which one can find a locally hyperbolic dispersion.  From
Eq.~(\ref{eq:8}), one can also define the critical angle $\theta_c$ from
which the dispersion is no longer hyperbolic:
\begin{equation}
\label{eq:5151}
\theta_c=\tan^{-1}\left[\frac{\sqrt{\Omega}}{ k_\mathrm{R}}\bigg/\sqrt{4-\frac{\Omega}{k_\mathrm{R}^2}}\right]\,.
\end{equation}
Corresponding to this hyperbolic region in momentum space, one can
then define a corresponding velocity range in real space. For each
point $(k_{x,i},k_{y,i})$ of the hyperbolic region limit---see the
green curve in Fig.~\ref{fig:3}(a)---we can derive a corresponding
velocity $(v_{x,i},v_{y,i})$ given by:
\begin{subequations}
\begin{align}
        v_{x,i}=\partial_{k_y} E(k_{x,i},k_y)\Bigr\rvert_{k_y=k_{y,i}}\,,\\
        v_{y,i}=\partial_{k_x} E(k_{x},k_{y,i})\Bigr\rvert_{k_x=k_{x,i}}\,.
\end{align}
\label{eq:9898}
\end{subequations}
Finally from $(v_{x,i},v_{y,i})$, we can then obtain a set of
coordinates defining a propagating distance
$(d_{x,i},d_{y,i})=(v_{x,i}t,v_{y,i}t)$. This set $(d_{x,i},d_{y,i})$
defines a closed surface in real space, that increases with time. This
area delimits the region of space into which self-interference can
occur. This is the 2D equivalent of the ``diffusion cone'' previously
derived in 1D~\cite{colas16a}. X-waves can thus be generated by
exciting $E_{-}(\boldsymbol{k})$ in this specific region, between two
inflection points $k_1$ and $k_2$, where the effective dispersion
appears locally hyperbolic.

The condensate dynamics can be obtained from a single-band
2D-Gross-Pitaevskii equation~\cite{khamehchi17a}:
\begin{equation}
\label{eq:9}
i\partial_t\psi(\boldsymbol{r})= \mathscr{F}^{-1}_{\boldsymbol{r}}[E_{-}(\boldsymbol{k})\psi(\boldsymbol{k})] +g_\mathrm{2D}|\psi(\boldsymbol{r})|^2\psi(\boldsymbol{r})\,,
\end{equation}
where $E_{-}(\boldsymbol{k})$ is the lower band defined in
Eq.~(\ref{eq:8}). $\mathscr{F}^{-1}_{\boldsymbol{r}}$ indicates the 2D
inverse Fourier transform, and $g_\mathrm{2D}$ the effective 2D
interaction strength.

The experiment of Khamehchi~\textit{et al.}~explored effectively one-dimensional dynamics, where the inital SOCBEC  was released from its initial cigar-shaped harmonic trap into a waveguide~\cite{khamehchi17a}.  The SOCBEC interaction energy was transformed into kinetic energy, leading to a spread in momentum space across the inflection point of the dispersion, and the development of a SIP~\cite{khamehchi17a,colas18a}.  Here we explore a similar scenario where a SOCBEC is released from a circularly symmetric harmonic trap into a two-dimensional waveguide, leading to the formation of a nonlinear X-wave.

As in the polariton case, only a weak nonlinearity is needed to
trigger the X-wave formation in a SOCBEC.  We choose
$g_{2D}N=7\times10^{-4}E_\mathrm{R}$, and an initial condensate size
of $\sigma_r=\unit{3.5}{\micro\meter}$, assumed to be Gaussian in this
regime~\cite{footnote1,pethick_book01a}. We impart an initial momentum to the wave packet of $(k_x^0,k_y^0) = (1.35,0)\times k_\mathrm{R}$
  which is within the inflection point region of the dispersion, as shown in
  Fig.~\ref{fig:3}(a,b). In Fig.~\ref{fig:3}(c--f) we present
selected time frames of the density evolution obtained from
Eq.~(\ref{eq:9}), along with the corresponding 1D-WT performed in the
direction of propagation at $y=0$, Fig.~(g--j). Once again, one can
observe the mechanism leading to the X-wave formation, that is, the
splitting of the wave packet into two sub-packets of different momenta
in a configuration where the faster packet is in a position to overlap
with the slower one and thus interfere with it. We
  note again the formation of vortex-antivortex pairs in
  Fig.~\ref{fig:3}(e), which have moved outside the boundary of Fig.~\ref{fig:3}(f).

We can perform a similar analysis for the energy of the system that we did for
the ideal hyperbolic case. The evolution of the different energy components is
presented in Fig.~\ref{fig:3}(k) and shows qualitatively the same features
as the hyperbolic case previously shown in
Fig.~\ref{fig:1}(k). One can, however, see that at $t=0$, the kinetic
energy is not zero, since the 2D-SOCBEC dispersion does not possess
the same $x$-$y$ symmetry.

For the parameters we have considered  the nonlinearity is strong enough to
form an X-wave, but remains weak enough to restrict the packet's
spread between the two inflection points $k_1$ and $k_2$. Increasing
the effective interaction strength would increase the packet's spread in
momentum and lead to the formation of more complex wave structures in
real space.

Without interactions ($g_{2D} = 0$) the internal reshaping of the wave packet does not occur, and the condensate dynamics are simply those of a slowly diffusing wave packet as shown in Fig.~\ref{fig:4}(a--d). However, in this case the SIP regime can still be reached by setting a tight Gaussian as initial condition~\cite{colas16a}.  Such dynamics are shown in Fig.~\ref{fig:4}(e--h). The real space density $|\psi(x,y)|^2$ displays self-interference fringes fully bounded in the delimiting area $d(x_i,y_i)$ previously derived (green line). In the $x$-$k$ space representation, the wavelet energy density closely follows the displacement associated with each wave vector $d(k_x,t)$. In the absence of interactions, the spread in momentum space is entirely defined from the initial condition through the wave packet's width $\sigma_\mathrm{r}$. 

Reaching the SIP regime requires a sufficiently broad wave packet in
momentum space that straddles the inflection points.  If the initial
wave packet does not have this structure, it can be achieved by a
transformation of interaction energy to kinetic
energy~\cite{colas18a}. To demonstrate this, we again take the
configuration used to generate the X-wave as in Fig.~\ref{fig:3}, but
with an interaction strength twice as large, $g_{2D} N=14\times10^{-4}E_\mathrm{R}$, shown in Fig.~\ref{fig:4}(i--l). At early times an X-wave
still forms thanks to the spread in momentum caused by the
nonlinearity, as shown in Fig.~\ref{fig:4}(i). The corresponding
wavelet transform shows the wave packet reshaping and the typical
feature of an X-wave self-interference, Fig.~\ref{fig:4}(k). However, at longer
times the X-wave shape in the density is no longer present and the
density exhibits a considerably more complex structure. The wavelet
analysis performed at this particular time of the evolution shows that
the packet's spread is now large enough to populate the dispersion
above the second inflection point, which is typical of the
SIP regime. This shows that X-waves generated in nonlinear systems only
exist and propagate for a finite time, and that more complicated effects can follow in their wake.

The internal reshaping of the wave packet due to a nonlinearity
leading to the X-wave formation is in many ways similar to the linear
self-interfering effect previously described for 1D
systems~\cite{colas16a, colas18a}. However the two mechanisms should
not be confused, even if they can both occur during the same
experiment, as shown in Fig.~\ref{fig:4}(i--l). The X-wave formation
mechanism exploits the spread in momentum space provided by the
nonlinear interaction to generate two distinct sub-packets, far from
the inflections points (if any) in the negative effective mass region, overlapping and interfering in real space. On
the other hand, the linear self-interference mechanism occurs due to
the change of sign of the $k$-dependent group velocity at the
inflection points to create an effective superposition across a
broad and continuous range of momenta.\\[2mm]

\section{Conclusions}
\label{sec:conclusions}
In this paper we have shown that nonlinear X-waves, including those
recently observed in excition-polariton systems, arise from an interference
mechanism triggered by the nonlinear interaction. The interaction
increases the packet's spread in momentum space, leading to the
formation of two effective sub-packets travelling at a different
velocities, hence overlapping in space and interfering. The complex
wave packet dynamics can be revealed and understood by utilising the
wavelet transform. The key ingredient in the X-wave formation is the
presence of a locally hyperbolic dispersion relation, and we have
shown that similar X-waves can be obtained in other physical systems
with this feature. For example, X-waves can be formed in
SOCBECs in the weakly interacting regime without the need for an
optical lattice potential.
Overall, our analysis of the X-wave formation dynamics utilising the wavelet transform provides
  physically insight into otherwise puzzling wave packet
  dynamics, and has identified the central role of self-interference.  This emphasizes the importance of the self-interfering packet effect for nonstandard
  dispersion relations either with or without the influence of nonlinearities.

The Supplemental Material for this manuscript includes movies of the
full dynamics for the three different systems we have considered in each of Figs.~\ref{fig:1},~\ref{fig:2},~\ref{fig:3} which shed further light on the nonlinear X-wave dynamics~\cite{footnote2}.

\begin{acknowledgements}
This research was supported by the Australian Research Council
  Centre of Excellence in Future Low-Energy Electronics Technologies
  (project number CE170100039) and funded by the Australian
  Government.  It was also supported by the Ministry of Science and
  Education of the Russian Federation through the Russian-Greek
  project RFMEFI61617X0085 and the Spanish MINECO under contract
  FIS2015-64951-R (CLAQUE).
\end{acknowledgements}

\begin{thebibliography}{36}
\bibitem[{foo({\natexlab{a}})}]{footnote3}
\bibinfo{note}{We set the initial condition with an imparted momentum of
  $k_x^0=1$ and not $k_x^0=0$. The dispersion relation is locally identical to
  the origin point, so the time evolved wavefunction density from this point
  would look the same, but without the extra linear momentum.}
\bibitem[{foo({\natexlab{e}})}]{footnote1}
\bibinfo{note}{For a $^{87}$Rb condensate this width would require a harmonic
  trap of frequency 150 Hz for g=0.}
\end{thebibliography}
\end{document}